\documentclass[aps,showpacs,twocolumn,draft,eqsecnum]{revtex4}
\usepackage{epsf}  

\begin{document}

\def\square
{\kern1pt\vbox{\hrule height 1.2pt\hbox{\vrule width1.2pt
\hskip 3pt\vbox{\vskip 6pt}\hskip 3pt\vrule width 0.6pt}
\hrule height 0.6pt}
\kern1pt}

\title{Fate of the first traversible wormhole: black-hole
collapse or inflationary expansion}

\author{Hisa-aki Shinkai}
\address{
Computational Science Division,
Institute of Physical \& Chemical Research (RIKEN),
Hirosawa 2-1, Wako, Saitama, 351-0198 Japan \\
\tt shinkai@atlas.riken.go.jp 
}

\author{Sean A. Hayward}
\address{Department of Science Education, Ewha Womans University, Seoul
120-750, Korea\\
\tt hayward@mm.ewha.ac.kr  
}

\date{10th May 2002 ~~ gr-qc/0205041}

\begin{abstract}
We study numerically the stability of Morris \& Thorne's first traversible
wormhole, shown previously by Ellis to be a solution for a massless ghost
Klein-Gordon field. Our code uses a dual-null formulation for spherically
symmetric space-time integration, and the numerical range covers both universes
connected by the wormhole. We observe that the wormhole is unstable against
Gaussian pulses in either exotic or normal massless Klein-Gordon fields. The
wormhole throat suffers a bifurcation of horizons and either explodes to form
an inflationary universe or collapses to a black hole, if the total input
energy is respectively negative or positive. As the perturbations become small
in total energy, there is evidence for critical solutions with a certain
black-hole mass or Hubble constant. The collapse time is related to the initial
energy with an apparently universal critical exponent. For normal matter, such
as a traveller traversing the wormhole, collapse to a black hole always
results. However, carefully balanced additional ghost radiation can maintain
the wormhole for a limited time. The black-hole formation from a traversible
wormhole confirms the recently proposed duality between them. The inflationary
case provides a mechanism for inflating, to macroscopic size, a Planck-sized
wormhole formed in space-time foam.
\end{abstract}

\maketitle
\section{Introduction}

Wormholes are known as a kind of solution of the Einstein equations, and have
become a popular research topic, raising theoretical possibilities of rapid
interstellar travel, time machines and warp drives. These topics sound like
science fiction, but after the influential study of traversible wormholes by
Morris \& Thorne \cite{MT}, it became widely accepted as a scientific topic
\cite{MVbook}. The only physically non-standard feature is that one has to
assume negative-energy matter to construct such a wormhole. However, such
exotic matter occurs extensively in quantum field theory and in alternative
gravitational theories such as scalar-tensor theories.

Until recently, wormholes have been studied mainly as static or cut-and-paste
models, or without an independently defined exotic matter model. Not long ago,
one of the authors \cite{wh} proposed a unified theory of black holes and
traversible wormholes, arguing that the two are dynamically interconvertible,
and that traversible wormholes are understandable as black holes under negative
energy density. This opens a new viewpoint on the dynamical nature of both
black holes and wormholes, including Hawking radiation. This synthesis has been
examined using a low-dimensional model \cite{HKL}, where the theory is
affirmed.

The purpose of this article is to investigate wormhole dynamics in
four-dimensional Einstein gravity, using numerical simulations. Our starting
point is a static wormhole which is perhaps best known as Morris \& Thorne's
opening example (their box 2). This metric is actually a solution for a
massless Klein-Gordon field whose gravitational coupling takes the opposite
sign to normal, as was shown earlier by several authors \cite{Cl}, the earliest
of which appears to be Ellis \cite{E}, who called it a drainhole. To our
knowledge, this is the earliest solution which would nowadays be called a
traversible wormhole.

We study dynamical perturbations of this static wormhole, using the spherically
symmetric Einstein system with the above exotic matter model, the massless
ghost Klein-Gordon field. We developed a numerical code based on a dual-null
coordinate system, in order to follow the horizon dynamics and radiation
propagation clearly. Our main experiment is to add or subtract Gaussian pulses
in the ghost field, i.e.\ respectively with negative or positive energy. We
also consider Gaussian pulses in a normal Klein-Gordon field, to see the effect
on the wormhole of normal matter, like a human being traversing the wormhole.
We discover how the initially static wormhole will change its structure due to
these dynamic perturbations. Although our model is specific to spherically
symmetric space-times, we believe that it illustrates the dynamical nature of
traversible wormholes. To our knowledge, this is the first numerical study of
wormhole dynamics.

We describe our model and numerical method in \S \ref{sec2} and present
numerical results in \S \ref{sec3}.  \S \ref{sec4} concludes.

\section{Model and Numerical method}\label{sec2}

\subsection{Field equations}
The field equations for a massless conventional and ghost Klein-Gordon field,
$\psi$ and $\phi$ respectively, in Einstein gravity are, in standard notation,
\begin{eqnarray}
&&R= 2\nabla\psi\otimes\nabla\psi-2\nabla\phi\otimes\nabla\phi, \\
&&\square \phi=0, \qquad \square \psi=0
\end{eqnarray}
where the coupling constants have been fixed by choice of units for $\psi$ and
$\phi$. We adopt a dual-null formulation developed by one of the authors
\cite{dne}. The spherically symmetric line-element may be written in dual-null
form
\begin{equation}
ds^2=r^2dS^2-2e^{-f}dx^+dx^-
\end{equation}
where $dS^2$ refers to the unit sphere. Then the configuration fields are
$(r,f,\phi,\psi)$ as functions of $(x^+,x^-)$. The Einstein equations in
spherical symmetry are expressed in \cite{sph,1st}, leading to the field
equations
\begin{eqnarray}
&& \partial_\pm\partial_\pm r +(\partial_\pm f)(\partial_\pm r)
  \nonumber \\
&&\qquad= -r(\partial_\pm\psi)^2 +r(\partial_\pm\phi)^2,\\
&&r^2\partial_+\partial_-f +2 (\partial_+r)(\partial_-r) +e^{-f}
\nonumber\\
&&\qquad= 2r^2(\partial_+\psi)(\partial_-\psi)
-2r^2(\partial_+\phi)(\partial_-\phi),
\\
&& r\partial_+\partial_-r+ (\partial_+r)(\partial_-r)+e^{-f}/2
=0,\\
&&r\partial_+\partial_-\phi +(\partial_+r)(\partial_-\phi)
+(\partial_-r)(\partial_+\phi)=0,
\\
&&r\partial_+\partial_-\psi +(\partial_+r)(\partial_-\psi)
+(\partial_-r)(\partial_+\psi)=0.
\end{eqnarray}
To obtain a system accurate near
$\Im^\pm$, we can use the conformal factor $\Omega=1/r$ instead of $r$ and, as
first-order variables, the conformally rescaled momenta
\begin{eqnarray}
\vartheta_\pm&=&2\partial_\pm r=-2\Omega^{-2}\partial_\pm\Omega,
\label{deftheta}\\
\nu_\pm&=&\partial_\pm f, \label{defnu} \\
\wp_\pm&=&r\partial_\pm\phi=\Omega^{-1}\partial_\pm\phi, \label{defwp}\\
\pi_\pm&=&r\partial_\pm\psi=\Omega^{-1}\partial_\pm\psi. \label{defpi}
\end{eqnarray}
Here $\vartheta_\pm$ has the meaning of expansion, though it is rescaled from
the usual definition, $\theta_\pm = 2 r^{-1} \partial_\pm r$. Then
$(\wp_\pm,\pi_\pm,\vartheta_\pm,\nu_\pm)$ are finite and generally non-zero at
$\Im^\mp$. The remaining first-order equations follow from the above field
equations and the spherically symmetric identity
$\partial_+\partial_-=\partial_-\partial_+$:
\begin{eqnarray}
\partial_\pm\vartheta_\pm&=&
-\nu_\pm\vartheta_\pm
-2\Omega\pi_\pm^2
+2\Omega\wp_\pm^2, \label{eq1}\\
\partial_\pm\vartheta_\mp&=&
-\Omega(\vartheta_+\vartheta_-/2+e^{-f}),\label{eq2}\\
\partial_\pm\nu_\mp&=&
-\Omega^2(\vartheta_+\vartheta_-/2+e^{-f} \nonumber \\
&&\qquad -2\pi_+\pi_-+2\wp_+\wp_-),
\label{eq3}\\
\partial_\pm\wp_\mp&=&-\Omega\vartheta_\mp\wp_\pm/2, \label{eq4}\\
\partial_\pm\pi_\mp&=&-\Omega\vartheta_\mp\pi_\pm/2. \label{eq5}
\end{eqnarray}
These equations plus the inverted momentum definitions
\begin{eqnarray}
\partial_\pm\Omega&=&-\Omega^2\vartheta_\pm/2,
\label{deftheta2}\\
\partial_\pm f&=&\nu_\pm, \label{defnu2} \\
\partial_\pm\phi&=&\Omega\wp_\pm, \label{defwp2}\\
\partial_\pm\psi&=&\Omega\pi_\pm \label{defpi2}
\end{eqnarray}
constitute the first-order dual-null form, suitable for numerical coding.

\subsection{Numerical method}
\begin{figure}[ht]
\setlength{\unitlength}{1mm}
\begin{picture}(80,80)
\put(0,0){\epsfxsize=70mm \epsffile{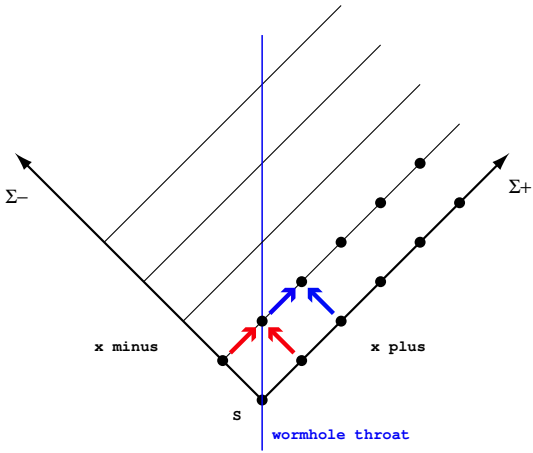} }
\end{picture}
\caption[quartic]{ Numerical grid structure.  Initial data are given on null
hypersurfaces $\Sigma_\pm$ ($x^\mp=0$, $x^\pm>0$) and their intersection $S$. }
\label{fig1_grid}
\end{figure}

We prepare our numerical integration range as drawn in Fig.\ref{fig1_grid}. The
grid will cover both universes connected by the wormhole throat $x^+=x^-$. The
grid is assumed to be equally spaced in $\Delta x^\pm=\Delta$.

The basic idea of numerical integration is as follows. We give initial data on
a surface $S$ and the two null hypersurfaces $\Sigma_\pm$ generated from it.
Generally the initial data have to be given as
\begin{eqnarray}
&&(\Omega,f,\vartheta_\pm,\phi,\psi)\quad\hbox{on $S$: $x^+=x^-=0$}\\
&&(\nu_\pm,\wp_\pm,\pi_\pm)\quad\hbox{on $\Sigma_\pm$: $x^\mp=0$, $x^\pm>0$.}
\end{eqnarray}
We then evolve the data $u=(\Omega, \vartheta_\pm, f, \nu_\pm, \phi,
\psi, \wp_\pm, \pi_\pm)$ on a constant-$x^-$ slice to the next.

Due to the dual-null decomposition, the causal region of a grid is clear, and
there are in-built accuracy checks: the integrability conditions or consistency
conditions $\partial_-\partial_+u=\partial_+\partial_-u$. In order to update a
point N (north), we have two routes from the points E (east) and W (west). The
set of equations, (\ref{eq1})-(\ref{defpi2}), gives us $x^+$-direction (W to N)
and $x^-$-direction (E to N) integrations together with the consistency
conditions.

All the equations can be written as
\begin{equation} u_{new}=u_{old}+ f(u_{RHS})\Delta,
\end{equation}
and the input variables, $u_{RHS}$, in the right-hand-side term depend on the
integration schemes. In more detail, we made the steps:
\begin{enumerate}
\item Integrate
$(\Omega, \vartheta_+, \vartheta_-, \nu_-, f, \phi, \psi, \wp_-, \pi_-)$ using
the $\partial_+$-equations (update N from W): use $u_{RHS}=u_{W}$.
\item Integrate
$(\Omega, \vartheta_+, \vartheta_-, \nu_+, f, \phi, \psi, \wp_+, \pi_+)$ using
the $\partial_-$-equations (update N from E): use $u_{RHS}=u_{E}$.
\item Update N from W again: use $u_{RHS}=(u_{W}+u_{N})/2$.
\item Update N from E again: use $u_{RHS}=(u_{E}+u_{N})/2$.
\item Check the consistency of
$(\Omega, \vartheta_+, \vartheta_-, f, \phi)$ at point N.  If the convergence
is not satisfactory, repeat back to the step 3. We set the tolerance as
$\max|u_{E \rightarrow N}-u_{W \rightarrow N}| < 10^{-5}$ for $(\Omega,
\vartheta_+, \vartheta_-, f, \phi)$.
\end{enumerate}
As a virtue of the dual-null scheme, we can follow the wormhole throat or
black-hole horizons easily. They are both trapping horizons, hypersurfaces
where $\vartheta_+=0$ or $\vartheta_-=0$ \cite{1st,bhd}. Another benefit is the
singular point excision technique. As we described, the causal region of each
grid point in the dual-null scheme is apparent.  When a grid point is inside a
black-hole horizon and near to the singularity, we can exclude that point and
grid points in its future null cone from further numerical computation.
Actually we excised grid points if one of
$(|\vartheta_+|,|\vartheta_-|,|e^{-f}|)$ exceeds $\sim 25$. We found that this
is useful for investigating inside a black-hole horizon, though there is
naturally a limit on how closely a singularity can be approached.

\subsection{Static wormhole}
The first Morris-Thorne wormhole metric \cite{MT} was given as
\begin{equation}
ds^2=(a^2+l^2)dS^2+dl^2-dt^2
\end{equation}
where $a$ denotes the throat radius of the wormhole. Since it is an overall
scale, we set $a=1$ for numerical purposes, but retain it in the text.
Transforming the proper radial length $l$ and the static proper time $t$ to
dual-null coordinates
\begin{equation}
x^\pm=(t\pm l)/\sqrt2
\end{equation}
the analytic expressions for the Ellis wormhole solution \cite{E} are
\begin{eqnarray}
\phi&=&\tan^{-1}(l/a),\\
\Omega&=&1/\sqrt{a^2+l^2},\\
f&=&0,\\
\wp_\pm&=&\pm a/\sqrt2\sqrt{a^2+l^2},\\
\vartheta_\pm&=&\pm \sqrt2l/\sqrt{a^2+l^2},\\
\nu_\pm&=&0 \label{EllisWH2}
\end{eqnarray}
with the conventional field $\psi$ vanishing. The wormhole is symmetric in
interchange of the two universes, $l\mapsto-l$, though this will not be so for
the perturbations. The energy density is
\begin{equation}
T_{tt}=-{a^2\over{8\pi(a^2+l^2)^2}}
\end{equation}
with units $G=1$. The negative energy density is a maximum at the throat $l=0$,
tends to zero at infinity and is smaller for larger wormholes. Note that the
wormhole throat is a double trapping horizon, $\vartheta_+=\vartheta_-=0$. The
behaviour of these generally different trapping horizons will be a key
indicator of the evolution of the perturbed wormhole.

The dual-null initial data for the static wormhole is
\begin{eqnarray}
&&(\phi,\Omega,f,\vartheta_\pm)=(0,1/a,0,0) \quad\hbox{on $S$} \\
&&(\wp_\pm,\nu_\pm)=(\pm a/\sqrt{2a^2+(x^\pm)^2},0)\quad\hbox{on $\Sigma_\pm$}.
\end{eqnarray}

\subsection{Gravitational mass-energy}
We find that the dynamical structure will be characterized by the total mass or
energy of the system, the Bondi energy. Localizing, the local gravitational
mass-energy is given by the Misner-Sharp energy $E$, while the (localized
Bondi) conformal flux vector components $\varphi^\pm$ were defined in
\cite{sph,1st}:
\begin{eqnarray}
E &=& (1/2) r [ 1 - g^{-1}(dr,dr)]
= (1/2)r + e^f r \, (\partial_+r)( \partial_- r)  \nonumber \\
&=& {1\over 2  \Omega} [1  + {1\over 2} e^f \, \vartheta_+  \vartheta_-]
\\
\varphi^\pm&=&r^2T^{\pm\pm}\partial_\pm r
= r^2e^{2f}T_{\mp\mp}\partial_\pm r \nonumber \\
&=&e^{2f}(\pi_\mp^2-\wp_\mp^2)\vartheta_\pm / 8 \pi.
\end{eqnarray}
They are related by the energy propagation equations or unified first law
\cite{1st}
\begin{equation}
\partial_\pm E = 4\pi\varphi_{\pm} = - {1\over 2} e^f(\pi_\pm^2-\wp_\pm^2)
\vartheta_\mp,
\end{equation}
so that one may integrate to
\begin{equation} E (x^+,x^-) = {a\over 2} +
4\pi\int_{(0,0)}^{(x^+,x^-)} (\varphi_+ dx^+ + \varphi_- dx^-),
\label{fluxintegrated}
\end{equation}
where the integral is independent of path, by conservation of energy
\cite{sph,1st}. Here the initial condition $E|_S=a/2$ corresponds to the static
wormhole. The definitions are local, but $\lim_{x^+\to\infty}E$ is the Bondi
energy and $\lim_{x^+\to\infty}\varphi_-$ the Bondi flux for the right-hand
universe. For the static wormhole, the energy $E=a^2/2\sqrt{a^2+l^2}$ is
everywhere positive, maximal at the throat and zero at infinity,
$l\to\pm\infty$, i.e.\ the Bondi energy is zero. Generally, the Bondi
energy-loss property, that it should be non-increasing for matter satisfying
the null energy condition \cite{sph}, is reversed for the ghost field.


\begin{figure}[t]
\setlength{\unitlength}{1mm}
\begin{picture}(85,50)
\put(0,0){\epsfxsize=80mm \epsffile{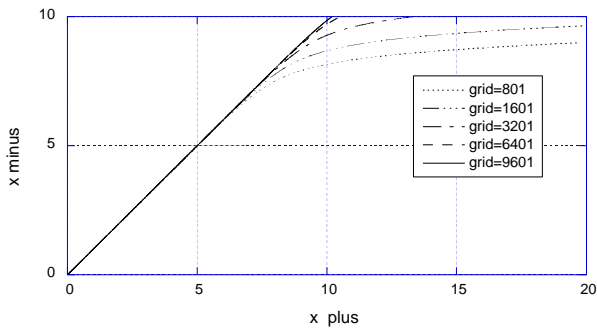} }
\end{picture}
\caption[quartic]{ Convergence behaviour of the code for exact static wormhole
initial data. The location of the trapping horizon $\vartheta_-=0$ is plotted
for several resolutions labelled by the number of grid
points for $x^+=[0,20]$. We see that
numerical truncation error eventually destroys the static configuration. }
\label{fig2_convergence}
\end{figure}

\begin{figure}[t]
\setlength{\unitlength}{1mm}
\begin{picture}(85,140)
\put(5,65){\epsfxsize=70mm \epsffile{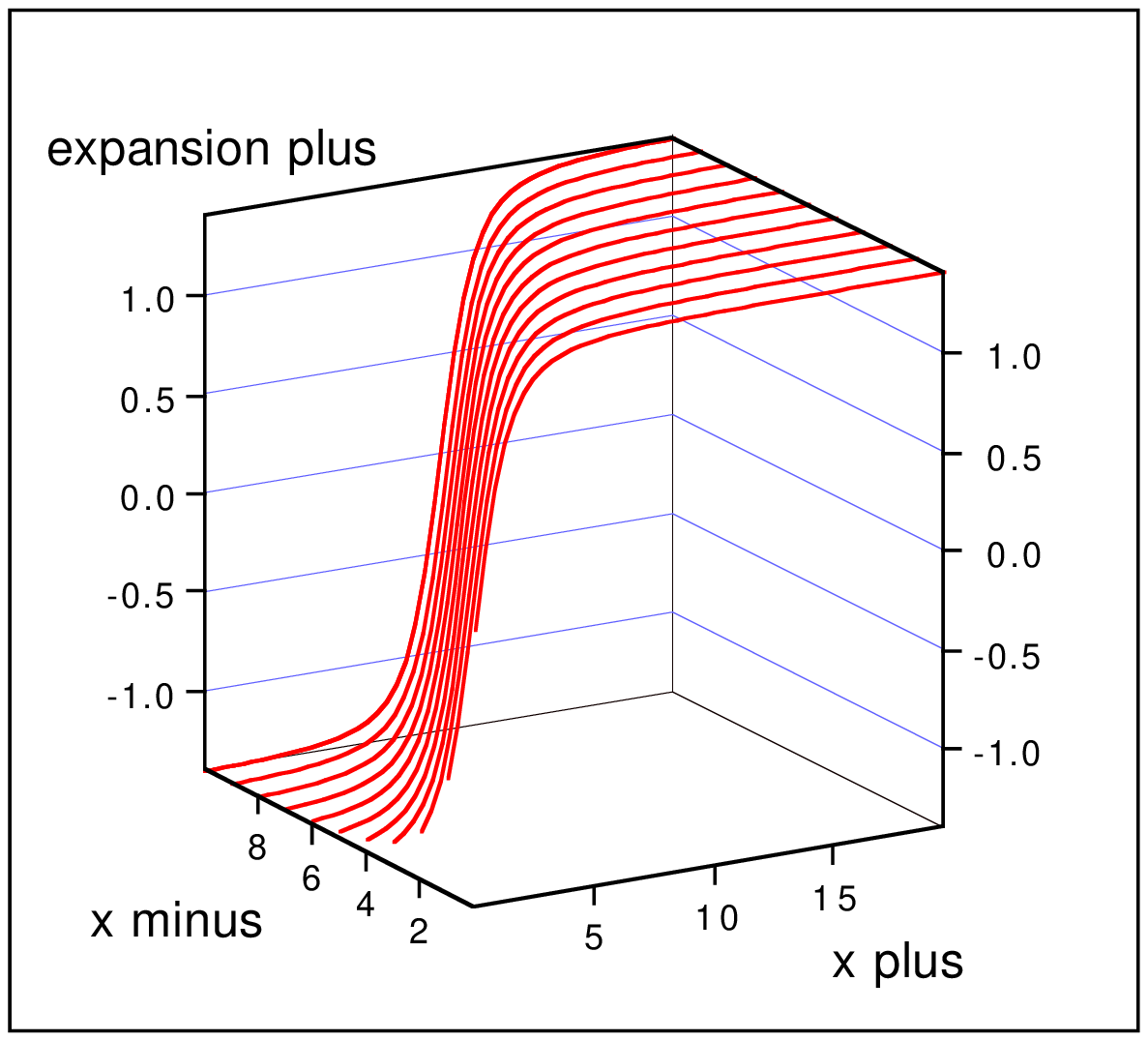} }
\put(5,0){\epsfxsize=70mm \epsffile{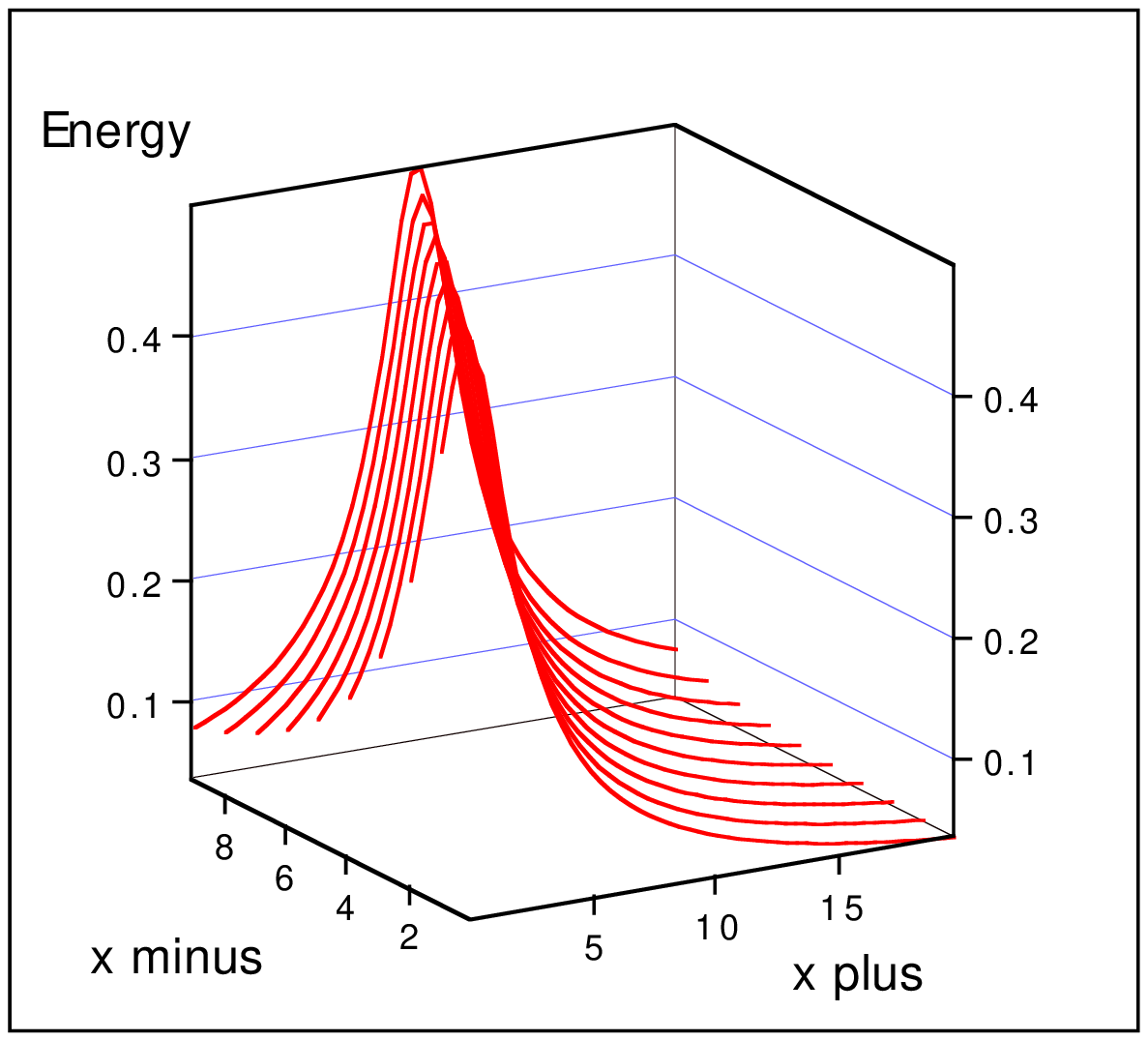} }
\end{picture}
\caption[quartic]{ Static wormhole configuration obtained with the highest
resolution calculation: (a) expansion $\vartheta_+$ and (b) local gravitational
mass-energy $E$ are plotted as functions of $(x^+, x^-)$. Note that the energy
is positive and tends to zero at infinity.
}
\label{fig3_config}
\end{figure}

\begin{figure}[h]
\setlength{\unitlength}{1mm}
\begin{picture}(85,70)
\put(10,0){\epsfxsize=60mm \epsffile{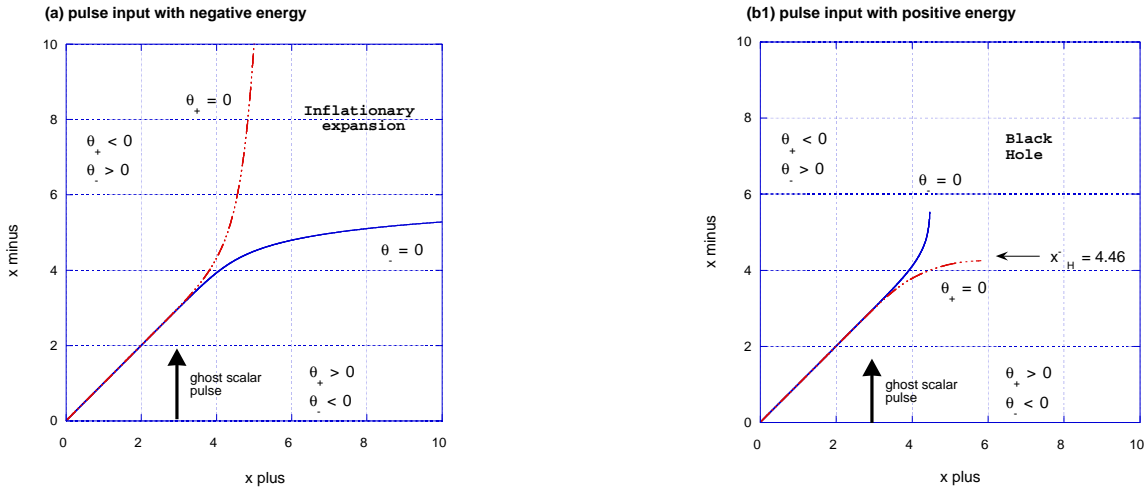} }
\end{picture}
\caption[quartic]{ Horizon locations, $\vartheta_\pm=0$, for perturbed
wormhole. Fig.(a) is the case we supplement the ghost field, $c_a = 0.1$, and
(b1) and (b2) are where we reduce the field, $c_a = -0.1$ and $-0.01$. Dashed
lines and solid lines are $\vartheta_+=0$ and $\vartheta_-=0$ respectively. In
all cases, the pulse hits the wormhole throat at $(x^+, x^-)=(3,3)$. A
$45^\circ$ counterclockwise rotation of the figure corresponds to a partial
Penrose diagram. } \label{fig4_pulseGexp}
\end{figure}

\setcounter{figure}{3}
\begin{figure}[h]
\setlength{\unitlength}{1mm}
\begin{picture}(85,140)
\put(10,70) {\epsfxsize=60mm \epsffile{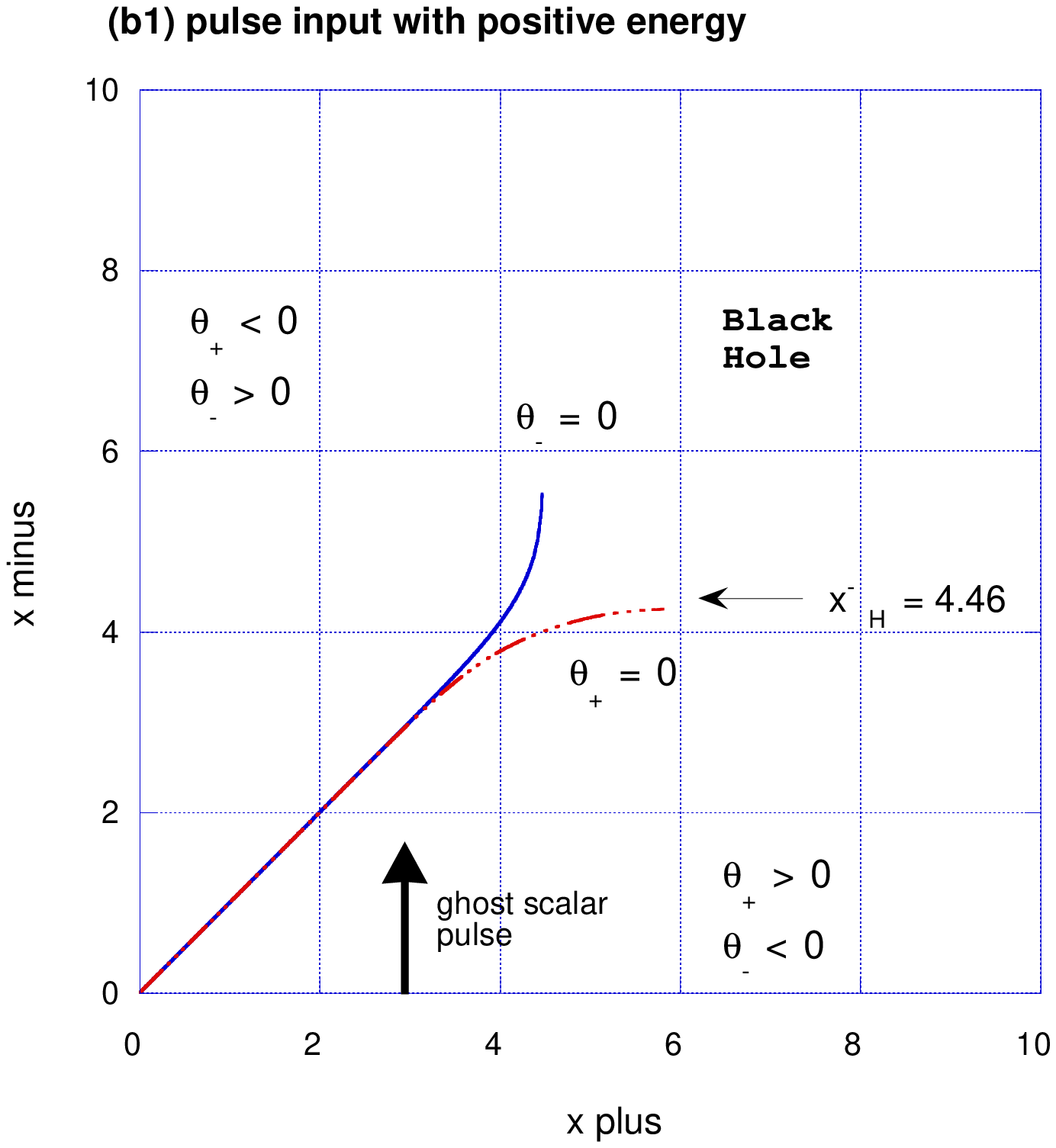} }
\put(10,0)  {\epsfxsize=60mm \epsffile{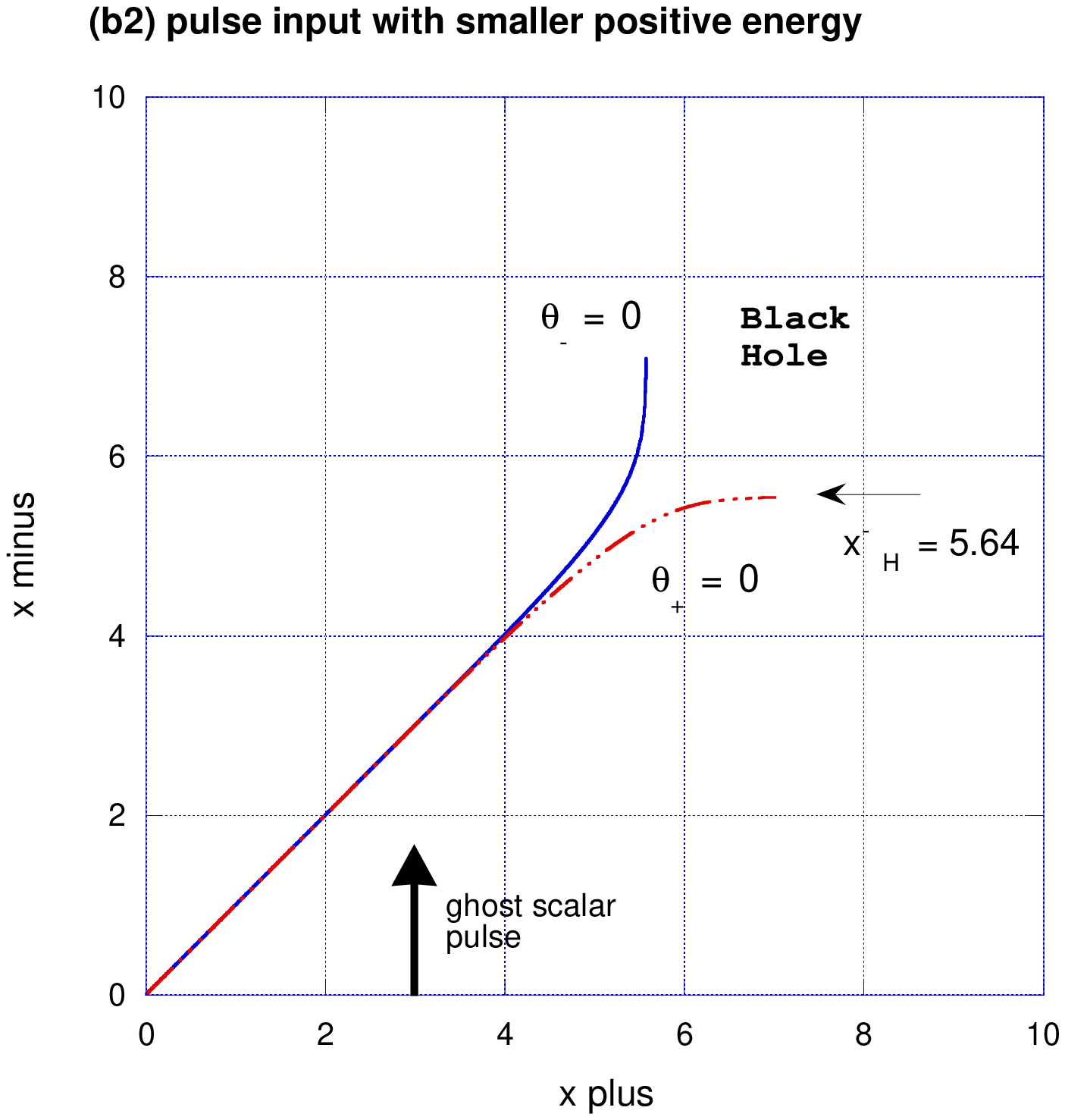} }
\end{picture}
\caption[quartic]{  (continued) }
\end{figure}

\begin{figure}[h]
\setlength{\unitlength}{1mm}
\begin{picture}(80,60)
\put(5,0){\epsfxsize=65mm \epsffile{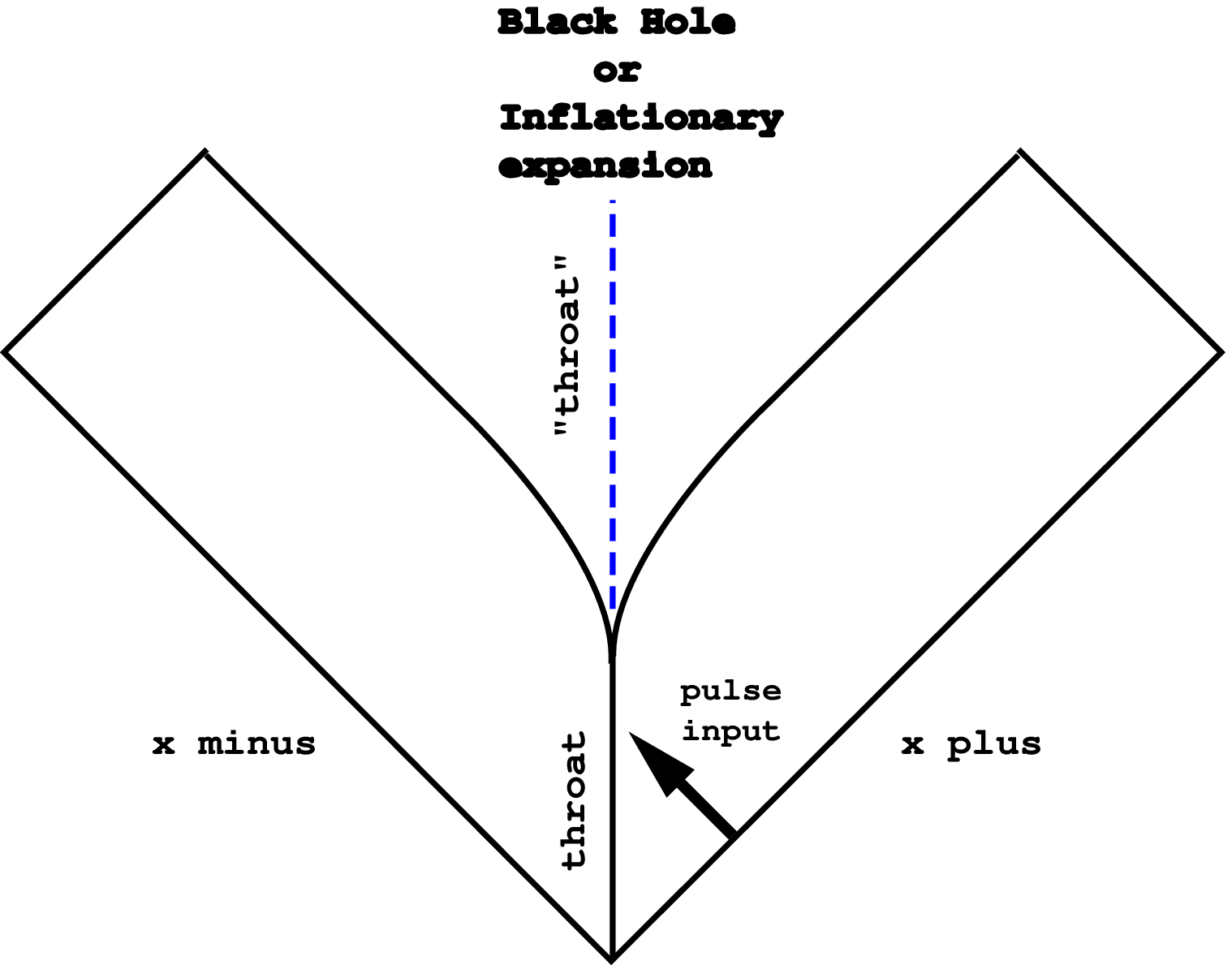} }
\end{picture}
\caption[quartic]{Partial Penrose diagram of the evolved space-time. }
\label{fig5_sketch}
\end{figure}

\begin{figure}[t]
\setlength{\unitlength}{1mm}
\begin{picture}(85,60)
\put(5,0){\epsfxsize=70mm \epsffile{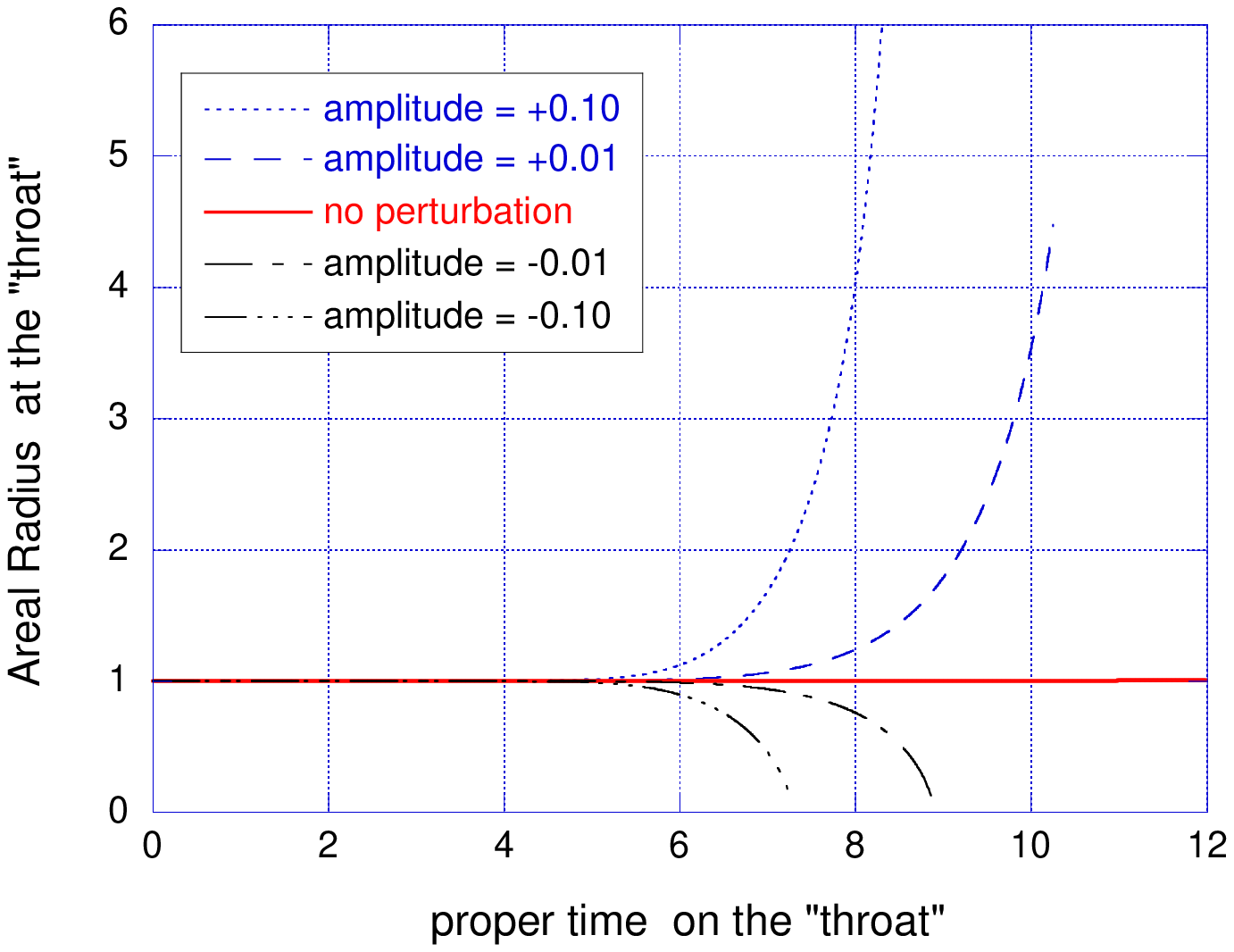} }
\end{picture}
\caption[quartic]{
Areal radius $r$ of the ``throat'' $x^+=x^-$, plotted as a function of proper
time. Additional negative energy causes inflationary expansion, while reduced
negative energy causes collapse to a black hole and central singularity.
  } \label{fig6_throat}
\end{figure}

\begin{figure}[t]
\setlength{\unitlength}{1mm}
\begin{picture}(85,160)
\put(5,110){\epsfxsize=65mm \epsffile{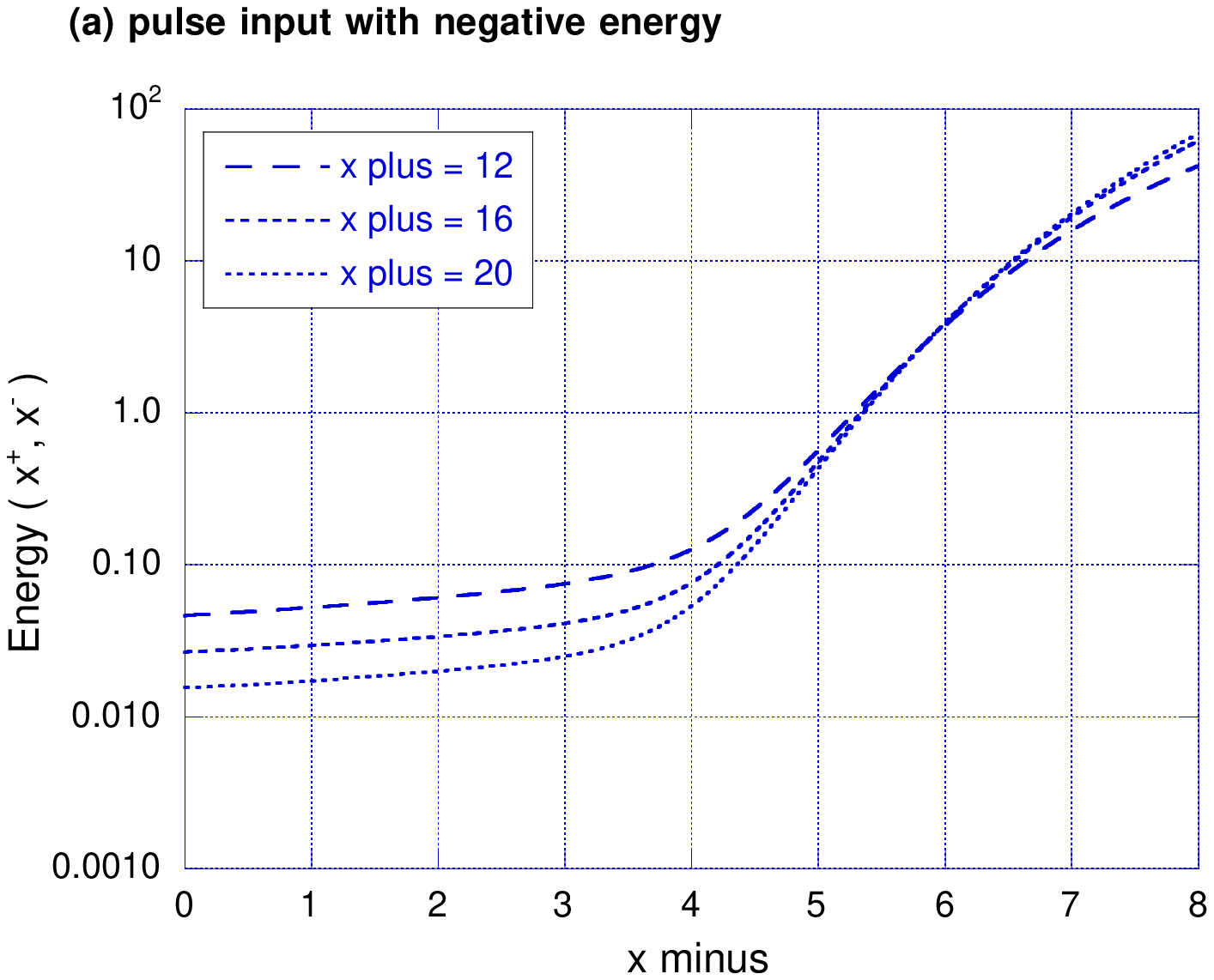} }
\put(5,55) {\epsfxsize=65mm \epsffile{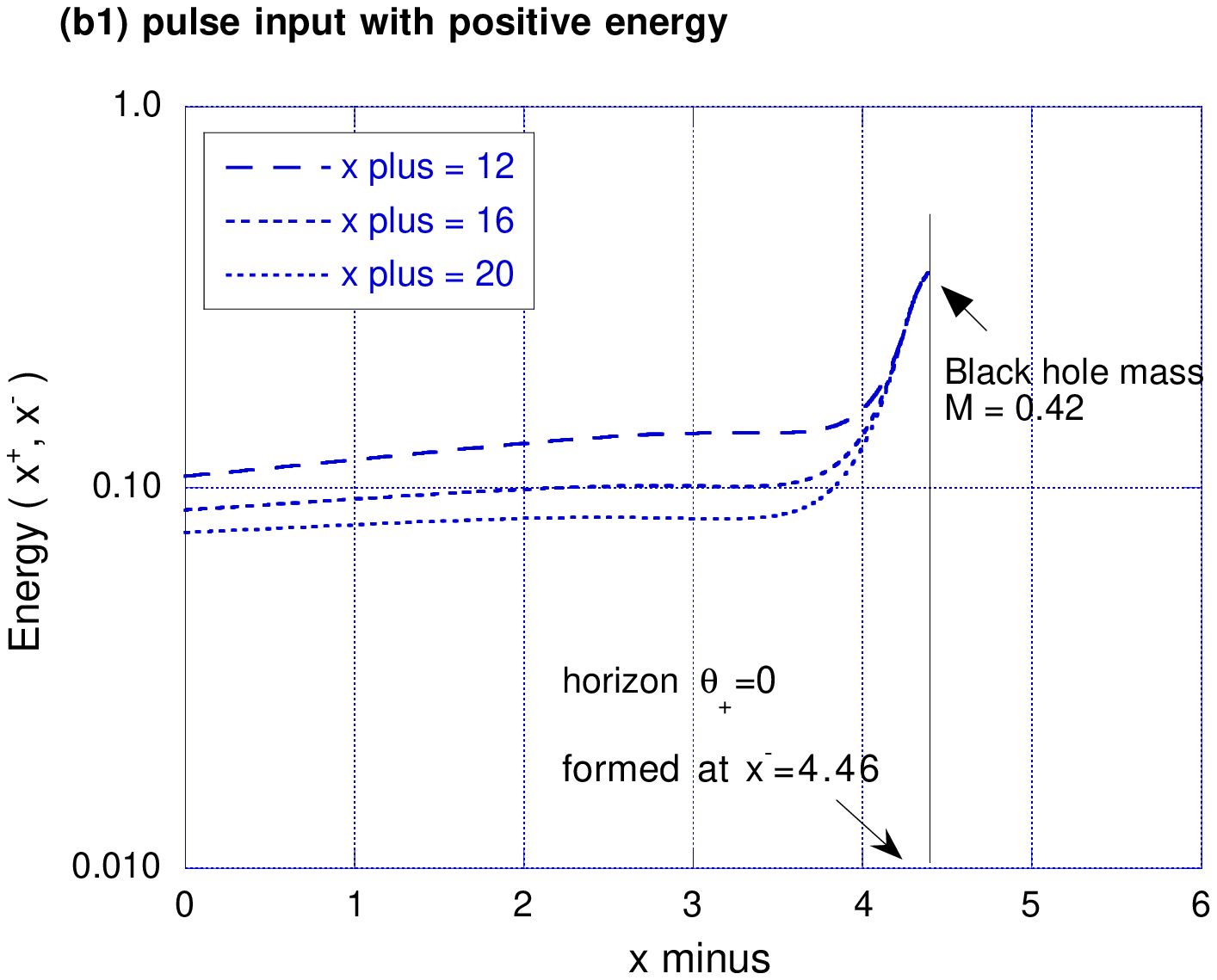} }
\put(5,0)  {\epsfxsize=65mm \epsffile{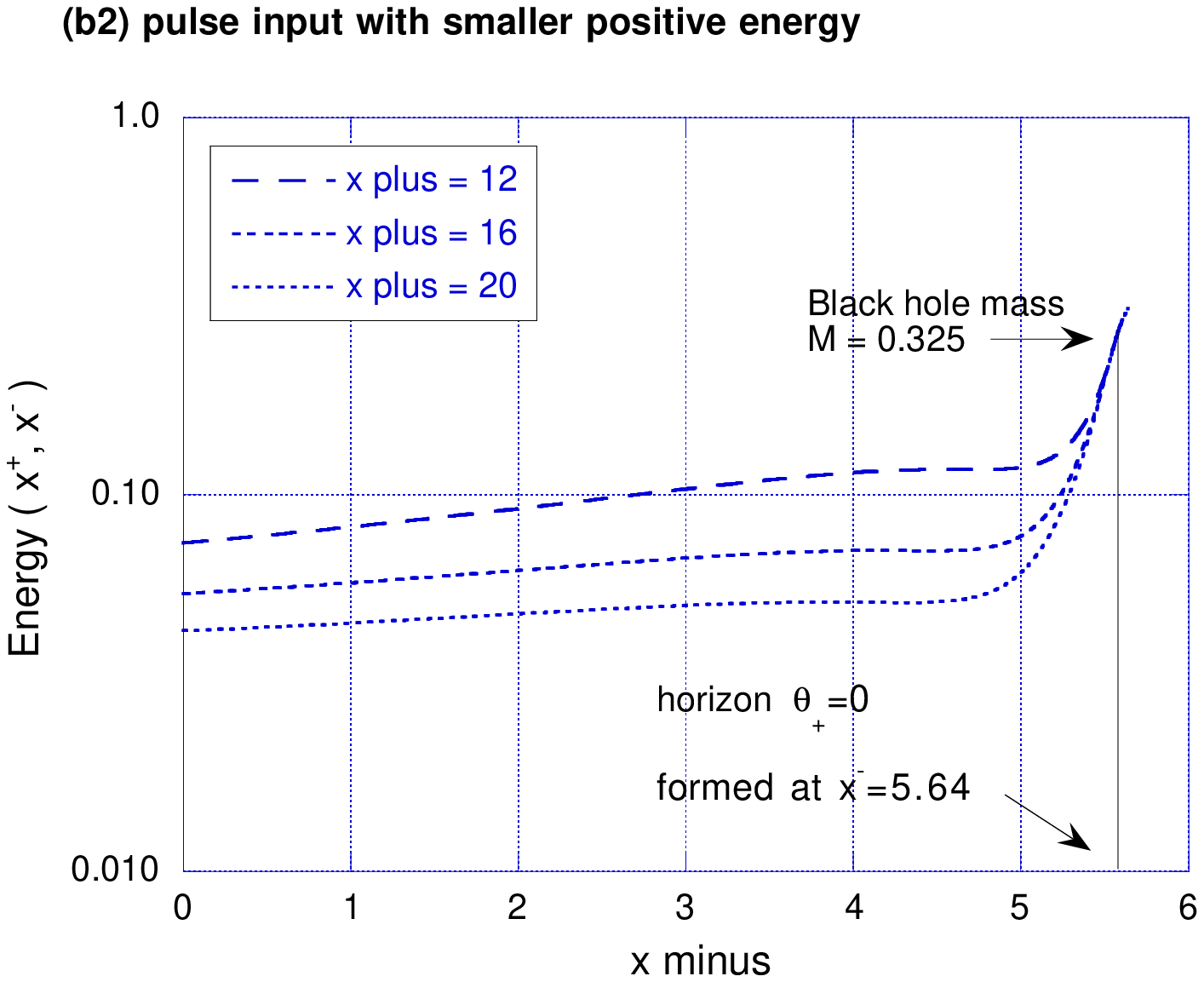} }
\end{picture}
\caption[quartic]{ Energy $E(x^+,x^-)$ as a function of $x^-$, for  $x^+ =
12,16,20$. Here $c_a$ is (a) $0.05$, (b1) $-0.1$ and (b2) $-0.01$. The energy
for different $x^+$ coincides at the final horizon location $x^-_H$, indicating
that the horizon quickly attains constant mass $M=E(\infty,x^-_H)$. This is the
final mass of the black hole. The values are shown in Table~\ref{table1}.
  } \label{fig7_localE}
\end{figure}

\begin{figure}[t]
\setlength{\unitlength}{1mm}
\begin{picture}(85,120)
\put(0,60){\epsfxsize=70mm \epsffile{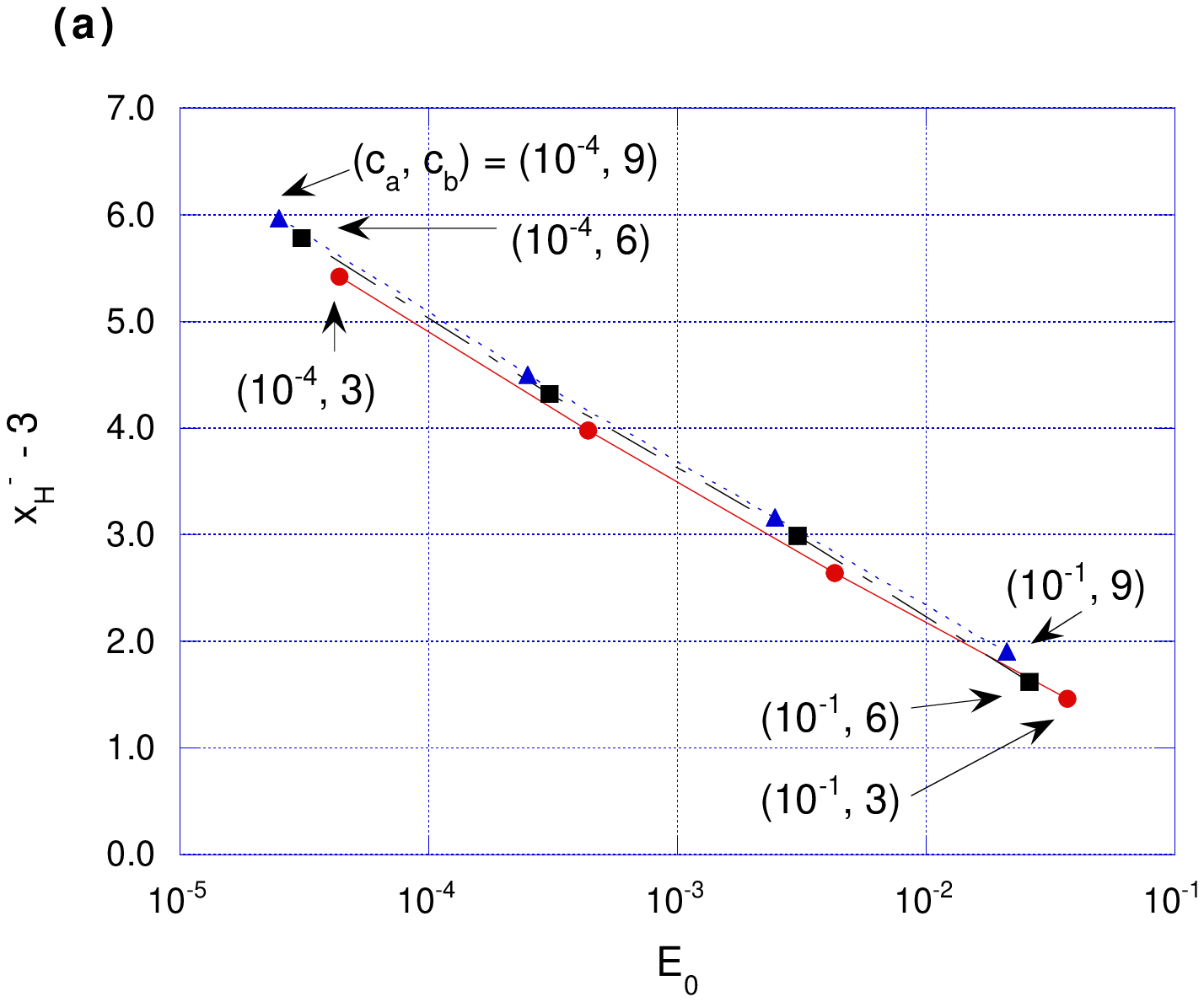} }
\put(0,0) {\epsfxsize=70mm \epsffile{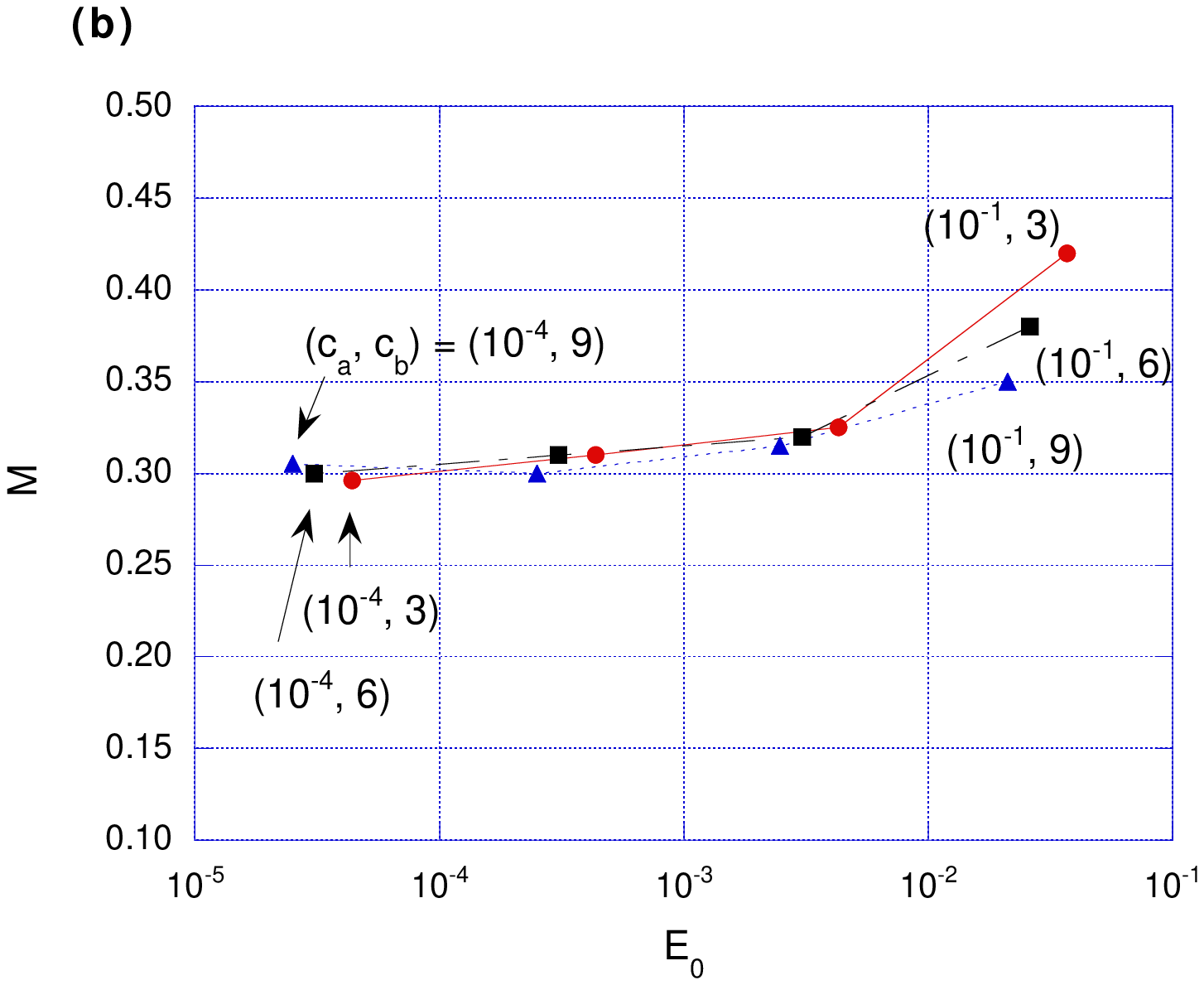} }
\end{picture}
\caption[quartic]{ Relation between the initial perturbation and the final mass
of the black hole. (a) The final horizon ($\vartheta_+=0$) coordinate
$x^-_H-3$, since we fixed $c_c=3$, versus initial energy of the perturbation,
$E_0$. We plotted the results of the runs of $c_a=10^{-1}, \cdots, 10^{-4}$
with $c_b=3, 6,$ and 9. They lie close to one line. (b) The final black hole
mass $M$ for the same examples. We see that $M$ appears to reach a non-zero
minimum for small perturbations.
  } \label{fig8_BHmass}
\end{figure}

\begin{figure}[bht]
\setlength{\unitlength}{1mm}
\begin{picture}(85,70)
\put(5,0) {\epsfxsize=60mm \epsffile{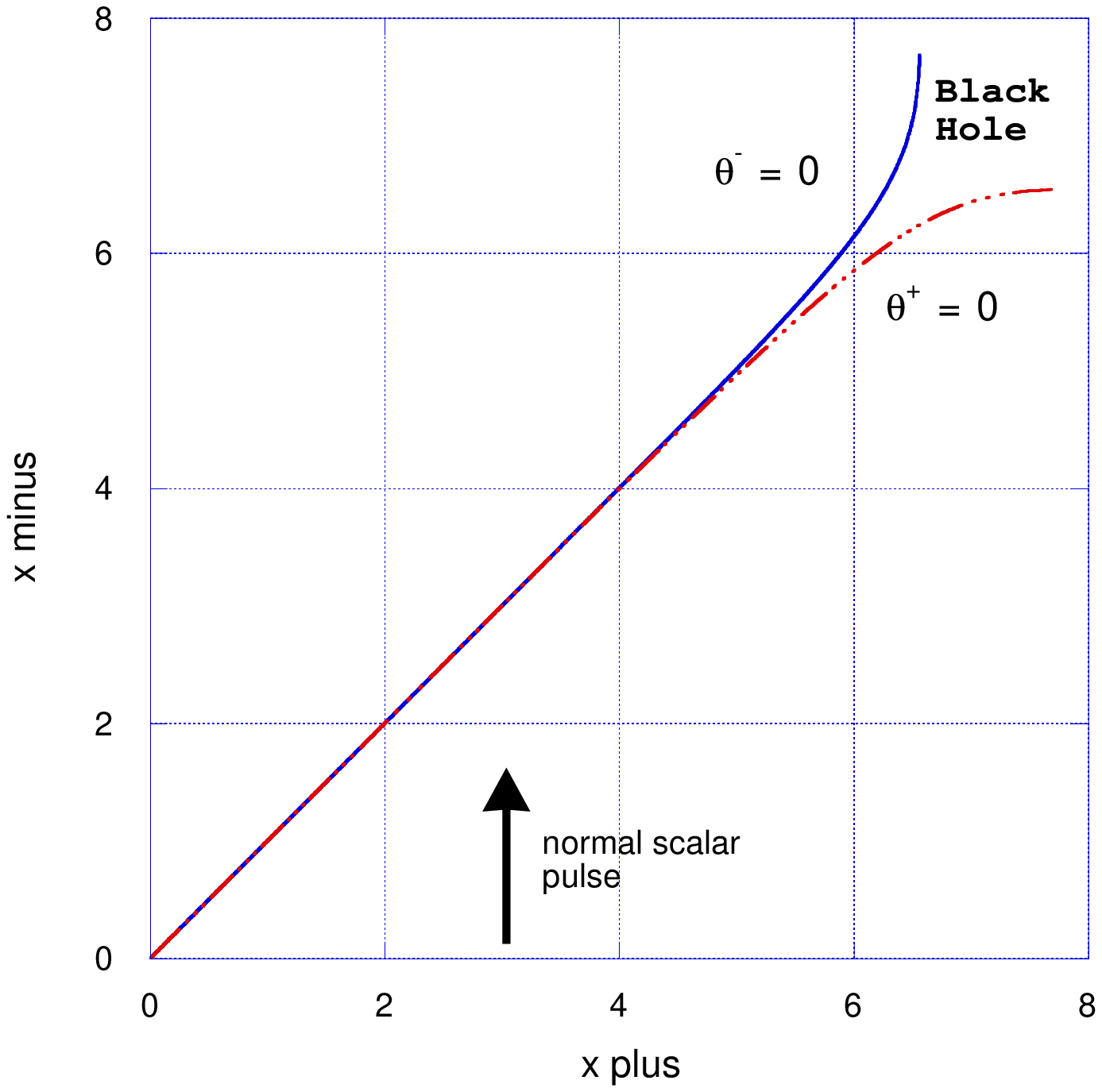} }
\end{picture}
\caption[quartic]{
Evolution of a wormhole perturbed by a normal scalar field.
Horizon locations:  dashed lines and solid lines are $\vartheta_+=0$
and $\vartheta_-=0$ respectively.
}
\label{fig9_pulseNexp}
\end{figure}

\begin{figure}[t]
\setlength{\unitlength}{1mm}
\begin{picture}(85,120)
\put(0,60){\epsfxsize=70mm \epsffile{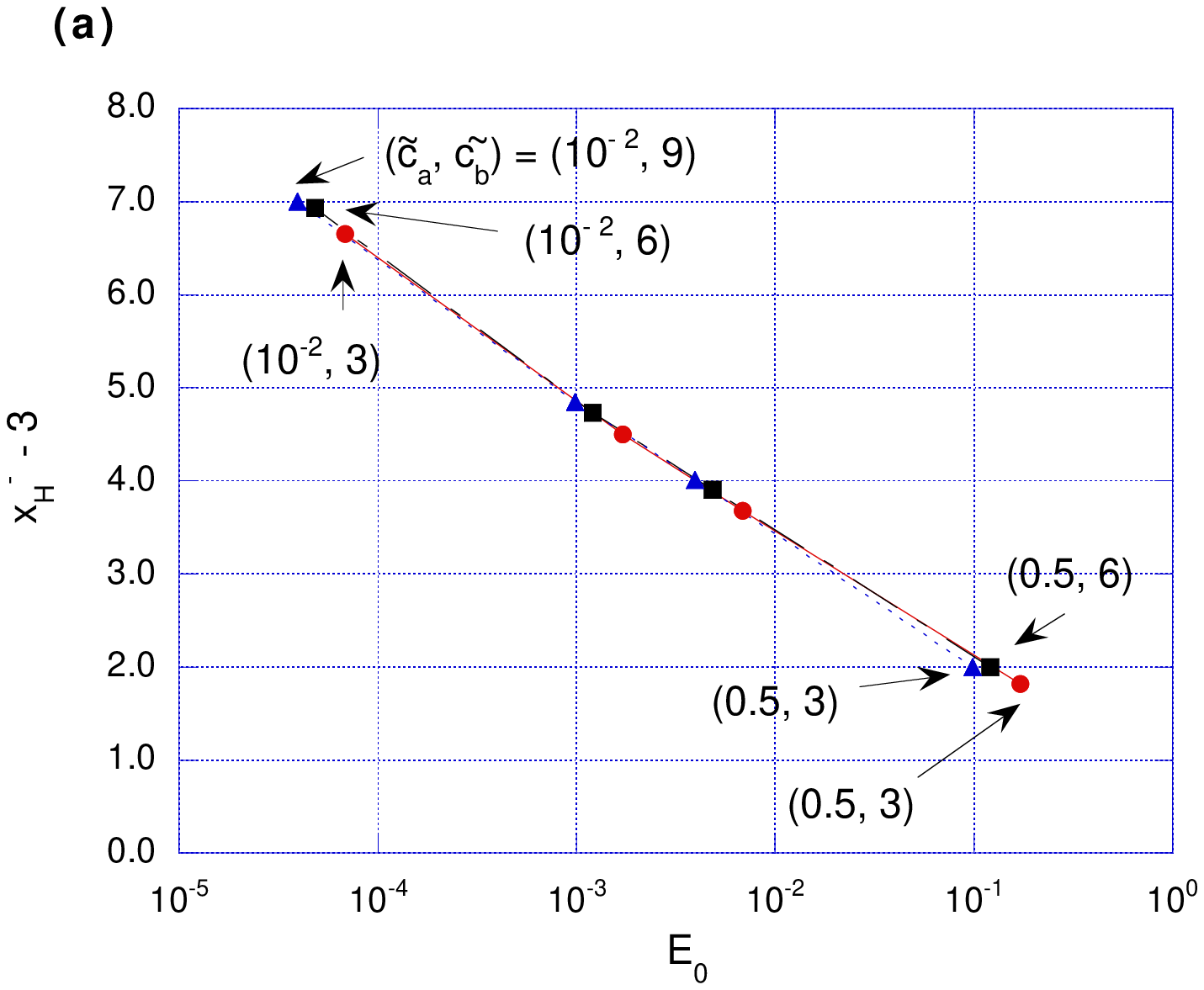} }
\put(0,0) {\epsfxsize=70mm \epsffile{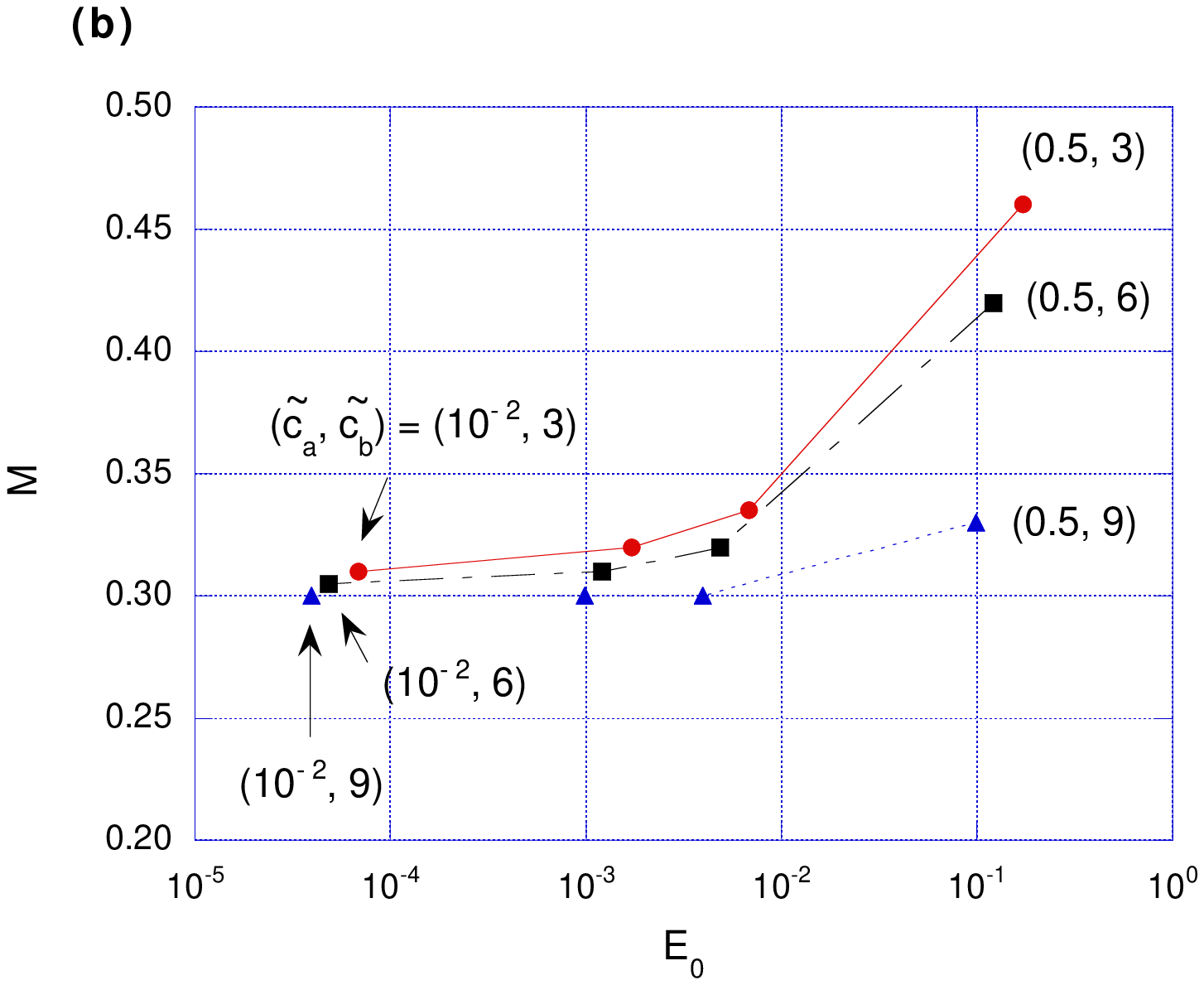} }
\end{picture}
\caption[quartic]{ The same plots as in Fig.\ref{fig8_BHmass} for conventional
field pulses. (a) The final horizon ($\vartheta_+=0$) coordinate $x^-_H-3$
versus initial energy of the perturbation, $E_0$. We plotted the results of the
runs of $\tilde{c}_a=0.5, \cdots, 10^{-2}$ with $\tilde{c}_b=3, 6,$ and 9. They
lie close to one line, different to that of Fig.\ref{fig8_BHmass}(a), but with
the same gradient. (b) The final black hole mass $M$ for the same examples.
  } \label{fig10_BHmass}
\end{figure}

\begin{figure}[bht]
\setlength{\unitlength}{1mm}
\begin{picture}(85,70)
\put(5,0) {\epsfxsize=60mm \epsffile{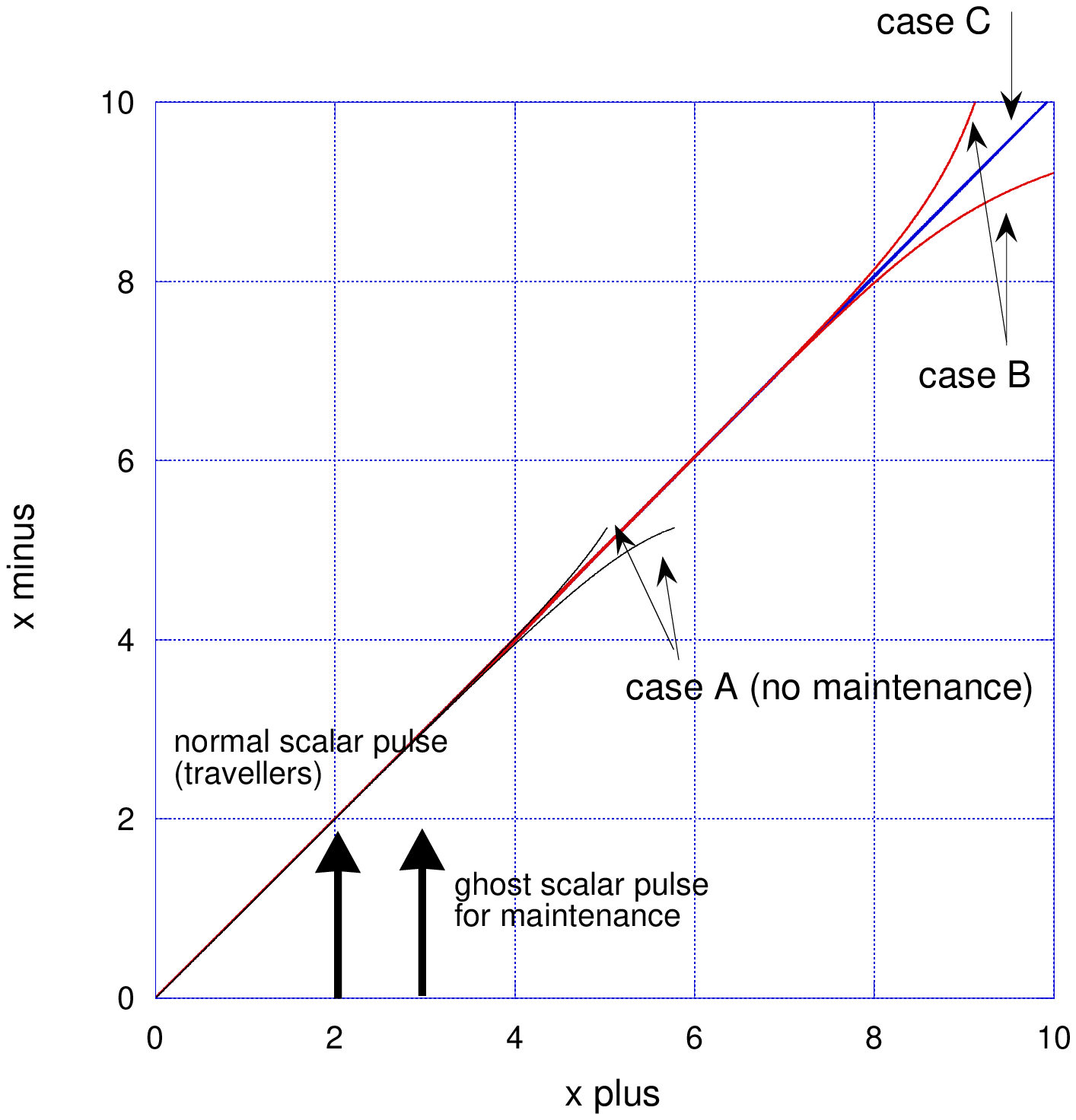} }
\end{picture}
\caption[quartic]{ Temporary wormhole maintenance.  After a normal scalar pulse
representing a traveller, we beamed in an additional ghost pulse to extend the
life of the wormhole. Horizon locations $\vartheta_+=0$ are plotted for three
cases: (A) no maintenance, which results in a black hole; (B) with a
maintenance pulse which results in an inflationary expansion; (C) with a more
finely tuned maintenance pulse, which keeps the static structure up to the end
of the range. } \label{fig11_maint}
\end{figure}


\section{Dynamical behaviour of traversible wormholes}\label{sec3}

\subsection{Evolution of static wormhole}
We start with the numerical evolution of the static wormhole. We find that
numerical truncation error can quite easily destroy the stability. In
Fig.\ref{fig2_convergence}, we show our tracking of the location of the
trapping horizon $\vartheta_-=0$ for static wormhole initial data, with
different numbers of grid points: 801, 1601, $\ldots$ 9601, for $x^+=[0,20]$.
The location of the throat, where $\vartheta_+=\vartheta_-=0$, is supposed to
be stationary at $x^+=x^-$.  The figure shows that at sufficient resolution, we
can maintain a static wormhole for a given time. Vice versa, this shows that in
our dual-null scheme and the current applied numerical integration scheme, the
lack of resolution inevitably causes an unstable evolution eventually. We
hereafter discuss the dynamical structure of perturbations of the wormhole only
in the reliable evolution range, with the resolution kept good enough (9601) to
preserve the static case.

In  Fig.\ref{fig3_config}, we demonstrate the numerical evolution of the static
wormhole, giving the expansion $\vartheta_+$ and the local gravitational
mass-energy $E$ as functions of $(x^+, x^-)$.

\subsection{Gaussian perturbations in ghost field}
We first put perturbations of the static wormhole in the form of Gaussian
pulses, input from the right-hand universe $l>0$. The initial data on
$\Sigma_+$ is
\begin{equation}
\wp_+=  a/\sqrt2\sqrt{a^2+l^2} + c_a \, \exp (-c_b(l-c_c)^2), \label{pulseeq}
\end{equation}
with all other initial data as for the static wormhole. Here $c_a,c_b,c_c$ are
parameters, with runs being performed for increasingly small values of the
amplitude $c_a$, $c_b=3,6,9$ and $c_c=3$. That is, the pulses will hit the
wormhole throat at $x^+=x^-=3$. For $c_a$ positive or negative, this
corresponds respectively to enhancing or reducing the supporting ghost field.
Note that this pulse-type perturbation starts only from the right-hand
universe. This situation is available because our numerical region covers both
sides of the wormhole, though we assume that the space-time is locally
spherically symmetric.

Fig.\ref{fig4_pulseGexp} shows the horizon locations for both $\vartheta_+=0$
and $\vartheta_-=0$, for $c_b=3$. We plotted three cases, $c_a = \pm 0.1$ and
$-0.01$, and it can be seen that in each case, the double trapping horizon
comprising the initial wormhole throat bifurcates when the pulse hits it. The
two horizons then accelerate away from each other, rapidly approaching the
speed of light and appearing to be asymptotically null. Due to the grid
excision technique for strong curvature regions, the final horizon coordinate
$x_H^-$, where the right-hand trapping horizon becomes null, is often excised
from the numerical computation.  In such a case, we determine $x_H^-$ by
extrapolating the horizon trajectory with a function $x^+=b_0+b_1 x^-+b_2
(x^-)^2+b_3/(x^- - x_H^-)$, where $b_i$ and  $x_H^-$ are unknown constants. A
partial Penrose diagram of the evolved space-time is given in
Fig.\ref{fig5_sketch}, which is like the prediction in \cite{wh}. However, the
two horizons move in opposite directions depending on the sign of $c_a$. In the
following, one should bear in mind that $\vartheta_+>0,\vartheta_-<0$ initially
in the right-hand universe $x^+>x^-$, with signs reversing for the left-hand
universe.

When we supplement the ghost field, $c_a = 0.1$, we see that the
$\vartheta_\pm=0$ horizons move respectively in the $x^\mp$ directions, meaning
that $\vartheta_\pm>0$ in the region between them. This defines past trapped
surfaces, which occur in expanding universes, such as the de~Sitter solution.
In Fig.\ref{fig6_throat} we plot the areal radius $r$ of the ``throat''
\footnote{Once the wormhole is perturbed, this hypersurface loses its specific
geometrical meaning as the wormhole throat $l=0$. However, one can say that it
is composed of minimal surfaces in some family of spatial hypersurfaces, which
are locally close to the constant-$t$ family for small perturbations.}
$x^+=x^-$, as a function of proper time
\begin{equation}
\tau = \int_{x^+=x^-} e^{-f/2}dt,\qquad t=(x^+ + x^-)/\sqrt{2}.
\end{equation}
It can be seen to rapidly increase, actually exponentially: the lines in the
figure can be expressed with a function $r/a=1+b_4 \exp(H (\tau -b_5))$ where
$b_i$ and $H$ are constants, and we show the Hubble constant $H$ in
Table~\ref{table1}. This exponential expansion, combined with the horizon
structure, indicates that the wormhole has exploded to an inflationary
universe. The two trapping horizons become the cosmological horizons. This
connection between wormholes and inflation was unanticipated, but is consistent
with the general theory \cite{wh}. The two universes connected by the wormhole
have essentially been combined into one universe.

On the other hand, when we reduce the ghost field, $c_a = -0.1$, the
$\vartheta_\pm=0$ horizons move respectively in the $x^\pm$ directions, meaning
that $\vartheta_\pm<0$ in the region between them. This defines future trapped
surfaces, which occur inside black holes, such as the Schwarzschild solution.
Fig.\ref{fig6_throat} shows that the areal radius is heading towards zero in
finite proper time, though the code cannot follow it into the apparent central
singularity. This and the horizon structure indicates that the wormhole has
collapsed to a black hole. The trapping horizons appear to become
asymptotically null, where they would coincide with the event horizons of the
black hole. This occurs around $x^-_H \sim 4.46$ for the right-hand
universe.

If the amplitude of the pulse is reduced, the black hole forms later, as shown
in Fig.\ref{fig4_pulseGexp}(b2) for $c_a=-0.01$. As a physical measure of the
size of the perturbation, we take the initial Bondi energy
\begin{equation}
E_0=E(\infty,0)
\end{equation}
scaled by the initial throat radius $a$. In practice we take $E(20,0)$ and
subtract the corresponding value for the static wormhole. Some data are
presented in Table \ref{table1}. We find that positive $E_0$ will cause
collapse to a black hole, while negative $E_0$ will cause explosion to an
inflationary universe. Also, the speed of the horizon bifurcation increases
with the energy. One could also scale the total energy by the maximal initial
energy $E$, but this occurs at the throat and is $a/2$. In either case, the
perturbations are small, down to 1\% in energy, yet the final structure is
dramatically different. Thus we conclude that the static wormhole is unstable.

Accepting that the static wormhole is stable to linear perturbations
\cite{BCCF,AP}, we seem to have discovered a non-linear instability, something
of interest in itself. Circumstantial evidence for a quadratic instability is
that it is first signalled by non-zero values of $\nu_\pm$, and the propagation
equations (\ref{eq3}) for $\nu_\pm$ have quadratic terms on the right-hand
side, which should cancel to zero in the static solution. This could be
addressed by second-order perturbation theory, but we know of no such studies
for wormholes.

In Fig.\ref{fig7_localE}, we plot the energy $E(x^+,x^-)$ as a function of
$x^-$, for  $x^+ = 12,16,20$ for the three cases in  Fig.\ref{fig4_pulseGexp}.
In each case, the mass increases rapidly in $x^-$ as the horizon
$\vartheta_\pm=0$ forms, but rapidly approaches a constant value in $x^+$ at
the horizon. This value
\begin{equation}
M=E(\infty,x^-_H)
\end{equation}
is the final mass of the black hole. Obtaining this easily is another virtue of
the dual null scheme. The values are shown in Table~\ref{table1}. The graphs
indicate that an observer at infinity will see a burst of radiation as the
wormhole collapses or explodes. For collapse, a certain fraction of the field
energy radiates away, the rest being captured by the black hole, constituting
its mass. For explosion, the radiated energy continues to rise in an apparently
self-supporting way as the universe inflates.

Fig.\ref{fig8_BHmass} gives a survey of various parameters of $c_a$ and $c_b$,
fixing $c_c=3$. Fig.\ref{fig8_BHmass}(a) shows how quickly the black-hole
horizon, the outer trapping horizon $\vartheta_+=0$, develops after the pulse
hits the throat. We see that the dots are close to one line, indicating a
logarithmic relation between the collapse time and the initial energy of the
perturbation,
\begin{equation}
e^{x^-_H-3} \propto E_0^{-0.60}.
\end{equation}
Fig.\ref{fig8_BHmass}(b) is the final black hole mass $M$ as a function of the
initial energy of the perturbation. Interestingly, as the perturbations become
small, the black-hole mass seems to approach a non-zero constant,
\begin{equation}
M\sim 0.30 a.
\end{equation}
In the inflationary case, the Hubble parameter $H$ (or horizon mass $1/2H$)
also seems to have a limiting value for small perturbations, as seen in
Table~\ref{table1}:
\begin{equation}
H\sim1.1/a.
\end{equation}
Since these masses are the same order as the wormhole scale $a$, we guess that
the static wormhole is something like an instanton with two different
attractors nearby. This scenario requires more detailed study with different
models, and we plan to report it subsequently.


\begin{table}[htb]
\begin{tabular}{c|c|c|l cc l}
\hline
  $c_a$ &  $c_b$ &   $E_0/a$  & \multicolumn{4}{l}{ final structure}
\\
\hline
   &   &   & &&{$Ha$}&
\\
\hline $+ 10^{-1}$ & 3.0 & $- 5.07 ~ 10^{-2}$ & \multicolumn{2}{l}{inflation}
&1.56 & (Fig.4a)
\\
$+ 10^{-2}$  & 3.0 &  $- 4.45 ~ 10^{-3}$ & \multicolumn{2}{l}{inflation}&1.19 &
\\
$+ 10^{-3}$  & 3.0 &  $- 4.40 ~ 10^{-4}$ & \multicolumn{2}{l}{inflation}&1.13 &
\\
$+ 10^{-4}$  & 3.0 &  $- 4.39 ~ 10^{-5}$ & \multicolumn{2}{l}{inflation}&1.12 &
\\
\hline
$ 0.0$   & 3.0 &  0.0
& \multicolumn{3}{l}{static}    &  (Fig.3)
\\
\hline $- 10^{-4}$  & 3.0 &  $+ 4.39 ~ 10^{-5}$  & black hole  & $8.42$ &
$0.296$  &
\\
$- 10^{-3}$  & 3.0 &  $+ 4.38 ~ 10^{-4}$  & black hole  & $6.98$ & $0.310$  &
\\
$- 10^{-2}$  & 3.0 &  $+ 4.32 ~ 10^{-3}$  & black hole  & $5.64$ & $0.325$  &
(Fig.4b2)
\\
$- 10^{-1}$  & 3.0 &  $+ 3.70 ~ 10^{-2}$  & black hole  & $4.46$ & $0.420$ &
(Fig.4b1)
\\
$- 10^{-1}$  & 6.0 &  $+ 2.60 ~ 10^{-2}$  & black hole  & $4.82$ & $0.38$ &
\\
$- 10^{-1}$  & 9.0 &  $+ 2.12 ~ 10^{-2}$  & black hole  & $4.90$ & $0.35$ &
\\
\hline \multicolumn{4}{l}{}& $x^-_H/a$ & $M/a$ &
\\
\hline
\end{tabular}
\caption{Initial Bondi energy $E_0$ and the final structures for different
ghost pulse inputs described by (\ref{pulseeq}). The energy is evaluated at
$x^+=20$, subtracting that of the exact static solution. The final structure
was judged at $x^-=10$ by the horizon structure. For inflation cases, the
Hubble constant $H$ was measured on the ``throat", Fig.\ref{fig5_sketch}. For
black-hole cases, we show the final horizon coordinate $x^-_H$, where the
$\vartheta_+=0$ trapping horizon becomes null, Fig.\ref{fig4_pulseGexp}(b), and
the final mass $M$ of the black hole, Fig.\ref{fig7_localE}(b).
  }\label{table1}
\end{table}

\begin{table}[tb]
\begin{tabular}{c|l|l|lccl}
\hline
  $\tilde{c}_a$ &  $\tilde{c}_b$ &   $E_0/a$  & 
  &$x^-_H/a$ &$M/a$ &
\\
\hline $\pm 1\times10^{-2}$  & 3.0 &  $+ 6.86 \times 10^{-5}$ & black hole &
$9.61$ & $0.31 $ &
\\
$\pm 5\times10^{-2}$  & 3.0 &  $+ 1.71 \times 10^{-3}$ & black hole & $ 7.50 $
& $0.32 $ &
\\
$\pm 1\times10^{-1}$ & 3.0 &  $+ 6.86 \times 10^{-3}$ & black hole & $6.68$ &
$0.34 $  & (Fig.9)
\\
$\pm 5\times10^{-1}$  & 3.0 &  $+ 1.71 \times 10^{-1}$ & black hole & $ 4.82 $
& $ 0.46 $ &
\\
$\pm 1\times10^{0}$  & 3.0 &  $+ 6.86 \times 10^{-1}$ & black hole & $3.70$ &
$0.90 $  &
\\
\hline
\end{tabular}
\caption{Initial Bondi energy $E_0$ and the final structures for different
conventional pulse inputs described by (\ref{pulseeq2}). The columns are the
same as for Table \ref{table2}. }\label{table2}
\end{table}


\subsection{Gaussian pulses in conventional field}
Similarly, we next consider adding a small amount of conventional scalar field
to the static wormhole solution. This is a simplified model of a situation that
a positive-energy creature jumps into a traversible wormhole. Again we add
Gaussian pulses to the initial data on $\Sigma_+$ as
\begin{equation}
\pi_+=  \tilde{c}_a \, \exp (-\tilde{c}_b(l-\tilde{c}_c)^2), \label{pulseeq2}
\end{equation}
with all other initial data as for the static wormhole. We set
$\tilde{c}_b=\tilde{c}_c=3$, so that the pulse will again hit the wormhole
throat at $x^+=x^-=3$.

We show the horizon structure in Fig.\ref{fig9_pulseNexp} and initial total
energy in Table \ref{table2}. The additional normal field slightly enhances the
total energy, with either sign of amplitude, and the wormhole collapses to a
black hole. The fundamental dynamical behaviour is the same as for the previous
ghost perturbations with positive energy. We see again a logarithmic relation
between the time and the initial energy of the perturbation, $e^{x^-_H-3} \propto
E_0^{-0.61}$, Fig.\ref{fig10_BHmass}(a). The critical exponent is consistent
with that found for perturbations in the ghost field, indicating a new kind of
critical behaviour reminiscent of that found by Choptuik \cite{Ch} in
black-hole collapse. In the limit of small perturbations, the black-hole mass
also appears to approach the same non-zero constant, $M\sim 0.30a$,
Fig.\ref{fig10_BHmass}(b). Although the current code has a limit on how small
perturbations can be, since the black-hole collapse should occur within the
reliable evolution range, the data suggest the existence of a critical solution
with a certain black-hole mass.

For these perturbations, the energy perturbation is as small as about 0.01\%, a
quite convincing indication of instability. From this and general principles
\cite{wh}, we expect that any normal matter traversing the wormhole will cause
its collapse to a black hole.

\subsection{Wormhole maintenance}
Supposing the normal field pulse represents an actual traveller of the
traversible wormhole, then he, she or it may go through the wormhole and exit
safely into the other universe if the speed is high enough, as can be seen from
the Penrose diagram obtained from Fig.\ref{fig9_pulseNexp}. However the
wormhole itself will collapse to a black hole, ending its usefulness for
travellers.

We here demonstrate a kind of temporary maintenance of the wormhole, by sending
in an additional ghost pulse just after the passing of the traveller. In
Fig.\ref{fig11_maint} we show the results for pulse parameters $(\tilde{c}_a,
\tilde{c}_b, \tilde{c}_c)=(0.1, 6.0, 2.0)$ for the normal field representing
the traveller, combined with a balancing pulse $(c_a, c_b, c_c)=(0.02390, 6.0,
3.0)$, case B, or $(c_a, c_b, c_c)=(0.02385, 6.0, 3.0)$, case C. If we do not
send the balancing pulse, the wormhole collapses to a black hole with horizons
given by case A in the plot. The case B ends up with an inflationary expansion,
while the case C keeps the wormhole structure at least until $x^-=10$. Since
the final fate of the wormhole is either a black hole or an inflationary
expansion, to keep the throat as it was requires a fine-tuning of the
parameters, and may not be realistic. However, it shows how the wormhole life
may be extended. This indicates that the wormhole might be maintained by
continual adjustments to the radiation level, though it would be a never-ending
project.

\section{Discussion}\label{sec4}
We have numerically studied the dynamic stability of the apparently earliest
\cite{E} and most frequently rediscovered \cite{wh,Cl} traversible wormhole,
most famous as the first Morris-Thorne wormhole \cite{MT}. We observe that the
wormhole is unstable against Gaussian pulses in both exotic and normal massless
Klein-Gordon fields. The wormhole throat suffers a bifurcation of horizons and
will either explode to an inflationary universe or collapse to a black hole, if
the total energy is respectively negative or positive. For the normal matter,
the total energy is necessarily positive and a black hole always results. This
black-hole formation from a traversible wormhole confirms the unified theory of
both \cite{wh}.

The inflationary expansion provides a mechanism for enlarging a quantum
wormhole to macroscopic size. Wheeler introduced wormholes to gravitational
physics as part of his still-popular idea of space-time foam \cite{W}, which
envisages transient wormholes at the Planck scale, due to quantum fluctuations
in space-time topology. Morris \& Thorne imagined pulling a wormhole out of
this foam and enlarging it somehow. Actual mechanisms are scarce: Roman
\cite{R} suggested inflation, though without an independently defined exotic
matter model, and Redmount \& Suen \cite{RS} found a quantum-mechanical
instability of certain cut-and-paste wormholes in a space-time foam model, also
leading to unbounded expansion. We have discovered that for one of the simplest
classical exotic matter fields, wormholes can naturally inflate. The problem is
only to stop the inflation proceeding indefinitely.

The results also provide an entertaining answer to the question of what happens
if someone attempts to traverse the wormhole. At best, with sufficient
alacrity, our hero could exit into the other universe, only to look behind and
see that the passage has caused the wormhole to collapse to a black hole,
thereby sealing off the causal connection to the home universe. Perhaps there
was also an unlucky companion who did not make it through, suffering the
well-known grisly fate of being trapped in a black hole. The survivor would be
left to ruminate on the ironic fate of those who would tamper with the fabric
of space-time: stranded forever in the twilight zone. As we showed, such a
tragedy could be avoided by a carefully calculated burst of ghost radiation,
just before or after the traveller passes, which maintains the static wormhole
for a certain time. The traveller then has a limited time to explore the other
universe before the wormhole collapses.

For our colleagues who are understandably sceptical about wormholes and exotic
matter, we hope that our numerical study has at least demonstrated that they
can be studied and understood by the same local, dynamical methods as used for
black holes with normal matter. Indeed, the results lead straight back to
mainstream physics of black holes and inflation, with new discoveries about
both, particularly the apparent critical behaviour. For a general audience,
perhaps the message is that (albeit theoretical) science can be stranger than
science fiction.

\begin{acknowledgments}
HS would like to thank Y.~Eriguchi, Y.~Kojima, T.~Harada and M.~Shibata for
comments, and S-W.~Kim for hospitality in visiting Ewha Womans University. SAH
would like to thank S-W.~Kim, S.~Sushkov, K.~Bronnikov and C.~Armendariz-Picon
for discussions. HS is supported by the special postdoctoral researchers
program at RIKEN, and this work was supported partially by the Grant-in-Aid for
Scientific Research Fund of Japan Society of the Promotion of Science,
No.~14740179. SAH is supported by Korea Research Foundation grant
KRF-2001-015-DP0095.
\end{acknowledgments}


\end{document}